\documentclass[12pt]{article}
\usepackage{epsfig}

\newcommand{\bea}{\begin{eqnarray}}
\newcommand{\eea}{\end{eqnarray}}
\newcommand{\be}{\begin{equation}}
\newcommand{\ee}{\end{equation}}

\newcommand{\as}{\alpha_s}
\newcommand{\asMZ}{\alpha_s(M^2_Z)}

\title{About QCD coupling constant \\ at NNLO from DIS data
\author{A.V.~Kotikov$^a$,  V.G.~Krivokhizhin$^a$, \\
B.G.~Shaikhatdenov$^a$, and G.~Parente$^b$ \\
\small $^a$ Joint Institute for Nuclear Research, Russia \\
\small $^b$ Universidade de Santiago de Compostela, Spain} }

\begin{document}
\maketitle
\abstract{ We give a brief review of recent QCD analysis~\cite{BKKP} carried out over
the deep inelastic scattering data on $F_2$ structure function and in the non-singlet approximation to the accuracy up to next-to-next-to-leading-order.
Specifically, analysis was performed over high statistics deep inelastic scattering data provided by BCDMS, SLAC, NMC and BFP collaborations.
For the coupling constant the following
value $\as(M_Z^2) = 0.1167 \pm 0.0022 
$
was found.
} \\

\section{ Introduction }

It goes without saying how it is crucial to know as accurate as possible the parton
distribution functions (PDFs) and the value of the strong coupling constant in order
to be able to make (relatively) solid predictions for various processes studied in a number of experiments. Within this realm, the deep inelastic scattering (DIS) of leptons off hadrons serves to be a cornerstone process to study PDFs which are universal and feed them further to other processes.

Nowadays the accuracy of data for DIS structure functions (SFs) makes it possible to study $Q^2$-dependence of logarithmic QCD-inspired corrections and those of power-like (non-perturbative) nature in a separate way (see for instance~\cite{Beneke} and references therein) which is important for the analysis to be performed according to a well defined scheme.

Until recently a commonly adopted benchmark tool for the analysis happened to be there
at the next-to-leading-order (NLO) level. However there have already appeared
papers in which QCD analysis of DIS SFs has been carried out up to the
next-to-next-to-leading order (NNLO) (see \cite{BKKP} and
\cite{PKK}--\cite{NNLOGluck}
and references therein).

In the present paper we show the results of our recent analysis~\cite{BKKP} of
of DIS SF $F_2(x,Q^2)$ with SLAC, NMC, BCDMS and BFP experimental
data involved~\cite{SLAC1}--\cite{BFP} at NNLO of massless perturbative QCD.

As in our previous paper the function $F_2(x,Q^2)$ is represented as a sum of the leading twist $F_2^{pQCD}(x,Q^2)$ and the twist four terms:
\be
F_2(x,Q^2)=F_2^{pQCD}(x,Q^2)\left(1+\frac{\tilde h_4(x)}{Q^2}\right)\,.
\label{1.1}
\ee

As is known there are at least two ways to perform QCD analysis over DIS data: the first one (see e.g.~\cite{ViMi,fits}) deals
with Dokshitzer-Gribov-Lipatov-Altarelli-Parisi (DGLAP) integro-differential equations~\cite{DGLAP} and let the data be examined directly, whereas the second one involves the SF moments and permits performing an analysis in analytic form as opposed to the former option.
In this work we take on the way in-between these two latter, i.e. analysis is carried out over the moments of SF $F_2^{k}(x,Q^2)$ defined as follows
\be
M_n^{pQCD/twist2/\ldots}(Q^2)=\int_0^1 x^{n-2}\,F_2^{pQCD/twist2/\ldots}(x,Q^2)\,dx
\label{1.a}
\ee
and then reconstruct SF for each $Q^2$ by using Jacobi polynomial expansion method \cite{Barker}-\cite{Kri1} (for further details see~\cite{KK2001}).
The theoretical input can be found
in the papers~\cite{KK2001,K2007}.

\section{ A fitting procedure }
\label{sec3}

To cut short this follows along the lines described in the
paper~\cite{KK2001}.
Here we just recall some aspects of the so-called polynomial expansion method.
The latter was first proposed in~\cite{Ynd} and further developed in~\cite{gon}. In these papers the method was based on the Bernstein polynomials and subsequently used to analyze data at NLO~\cite{KaKoYaF,KaKo} and NNLO level~\cite{SaYnd,KPS1}.
The Jacobi polynomials for that purpose were first proposed and then subsequently developed in~\cite{Barker,Kri,Kri1} and used in~\cite{PKK}-\cite{KPS1},~\cite{Vovk}.

With the QCD expressions for the Mellin moments $M_n^{k}(Q^2)$ analytically calculated according to
the formul\ae\, given above the SF $F_2^k(x,Q^2)$ is reconstructed by using the Jacobi polynomial expansion method:
$$
F_{2}^k(x,Q^2)=x^a(1-x)^b\sum_{n=0}^{N_{max}}\Theta_n ^{a,b}(x)\sum_{j=0}^{n}c_j^{(n)}(\alpha ,\beta )
M_{j+2}^k (Q^2)\,,
\label{2.1}
$$
where $\Theta_n^{a,b}$ are the Jacobi polynomials and $a,b$ are the parameters fitted. A condition
put on the latter is the requirement of the error minimization while reconstructing the structure functions.

Since a twist expansion starts to be applicable only above $Q^2 \sim 1$ GeV$^2$
the cut $Q^2 \geq 1$ GeV$^2$ on data is applied throughout.

MINUIT program~\cite{MINUIT} is used to minimize two variables
$$
\chi^2_{SF} = \biggl|\frac{F_2^{exp} - F_2^{teor}}{\Delta F_2^{exp}}\biggr|^2\,, \qquad
\chi^2_{slope} = \biggl|\frac{D^{exp} - D^{teor}}{\Delta D^{exp}}\biggr|^2\,,
$$
where $D=d\ln F_2/d\ln\ln Q^2$. Quality of the fits is characterized by
$\chi^2/DOF$ for the SF
$F_2$. Analysis is also performed for the SF slope $D$ that serves the purpose of checking the properties of fits.

We use free normalizations of the data for different experiments.
For a reference set, the most stable deuterium BCDMS data at the value of the
beam initial energy $E_0=200$ GeV is used.

\section{Results}

In the nonsinglet approximation gluons are not taken into account,
hence the cut on Bjorken variable ($x\geq 0.25$) imposed
where the gluon density is believed to be negligible.

The starting point of the evolution is taken to be
$Q^2_0$ = 90 GeV$^2$.
This latter value is close to the average value of $Q^2$ spanning the corresponding data.
On grounds of previous knowledge the maximal value of the number
of moments to be accounted for is $N_{max} =8$~\cite{Kri,Kri1} (though we
check $N_{max}$ dependence just like in the NLO analysis) and
the cut $0.25 \leq x \leq 0.8$ is imposed everywhere.


Analogously to what has been done in~\cite{KK2001,BKKP}
the cut on the Bjorken variable $x$ is imposed in combination with those placed on the $y$ variable as follows:
\bea
& &y \geq 0.14 \,~~~\mbox{ for  }~~~ 0.3 < x \leq 0.4 \nonumber \\
& &y \geq 0.16 \,~~~\mbox{ for  }~~~ 0.4 < x \leq 0.5 \nonumber \\
& &y \geq 0.23 ~~~\mbox{ for  }~~~ 0.5 < x \leq 0.6 \nonumber \\
& &y \geq 0.24 ~~~\mbox{ for  }~~~ 0.6 < x \leq 0.7 \nonumber \\
& &y \geq 0.25 ~~~\mbox{ for  }~~~ 0.7 < x \leq 0.8\,, \nonumber
\eea
which are meant to cut out those points suffering large systematic errors.
Then a full set of data to be analyzed consists of 797 points.

To verify a range of applicability of perturbative QCD
we start with analyzing the data without a contribution coming from the twist-four terms (which means $F_2 = F_2^{pQCD}$) and perform several fits with the cut
$Q^2 \geq Q^2_{min}$ gradually increased.
From Table~1 it is seen that unlike the NLO analysis
quality of the fits starts to appear fairly good from $Q^2=7$ GeV$^2$ onwards.
For the purpose of comparison, the twist-four corrections are added and the data with a general cut $Q^2 \geq 1$ GeV$^2$ imposed upon is fitted. It is clearly seen that as in the NLO case here the higher twists do sizably improve the quality of the fit, with insignificant discrepancy in the values of the coupling constant to be quoted below.

\vspace{0.5cm}
{\bf Table 1.} $\asMZ$ and $\chi^2$ in the case of the combined analysis
\begin{center}
\begin{tabular}{|l|c|c|c|c|c|c|}
\hline
& &  & &  &\\
$Q^2_{min}$ & $N$ of & HTC &$\chi^2(F_2)$/DOF &
$\as(90~\mbox{GeV}^2)$ $\pm$ stat & $\asMZ$ \\
& points &  &  & &  \\
\hline \hline
1.0 & 797 &  No & 2.20 & 0.1767 $\pm$ 0.0008 & 0.1164 \\
2.0 & 772 &  No & 1.14 & 0.1760 $\pm$ 0.0007 & 0.1162 \\
3.0 & 745 &  No & 0.97 & 0.1788 $\pm$ 0.0008 & 0.1173 \\
4.0 & 723 &  No & 0.92 & 0.1789 $\pm$ 0.0009 & 0.1174 \\
5.0 & 703 &  No & 0.92 & 0.1793 $\pm$ 0.0010 & 0.1176 \\
6.0 & 677 &  No & 0.92 & 0.1793 $\pm$ 0.0012 & 0.1176 \\
7.0 & 650 &  No & 0.92 & 0.1782 $\pm$ 0.0015 & 0.1171 \\
8.0 & 632 &  No & 0.93 & 0.1773 $\pm$ 0.0018 & 0.1167 \\
9.0 & 613 &  No & 0.93 & 0.1764 $\pm$ 0.0022 & 0.1163 \\
10.0 & 602 & No & 0.92 & 0.1742 $\pm$ 0.0023 & 0.1154 \\
11.0 & 588 & No & 0.91 & 0.1718 $\pm$ 0.0027 & 0.1144 \\
12.0 & 574 & No & 0.92 & 0.1717 $\pm$ 0.0029 & 0.1143 \\
13.0 & 570 & No & 0.92 & 0.1710 $\pm$ 0.0030 & 0.1140 \\
14.0 & 562 & No & 0.92 & 0.1712 $\pm$ 0.0032 & 0.1141 \\
15.0 & 550 & No & 0.91 & 0.1715 $\pm$ 0.0033 & 0.1142 \\
\hline \hline
1.0 & 797 & Yes & 0.98 & 0.1772 $\pm$ 0.0027 & 0.1167 \\
\hline
\end{tabular}
\end{center}
\vspace{0.5cm}

The following values for parameters in the parametrizations of the parton distributions for the case corresponding to the last row of the above table are obtained:
\bea
A^{H_2}_{NS} &=& 2.54,~~~\, A^{D_2}_{NS} ~=~ 2.38,~~~ A^C_{NS} ~=~ 3.29,
~~~ A^{Fe}_{NS} ~=~ 2.35,  \nonumber \\
b^{H_2}_{NS} &=& 4.16,~~~\, b^{D_2}_{NS} ~=~ 4.22,~~~\,\, b^C_{NS} ~=~ 4.23,~~~\,\,
b^{Fe}_{NS} ~=~ 4.39,   \nonumber \\
d^{H_2}_{NS} &=& 6.08,~~~\, d^{D_2}_{NS} ~=~ 3.89,~~~\, d^C_{NS} ~=~ 2.02,~~~\,\,
d^{Fe}_{NS} ~=~ 3.31. \nonumber
\eea

The parameter values of the twist-four term are presented in Table~2.
Note that these for $H_2$ and $D_2$ targets are obtained in separate fits by analyzing SLAC, NMC and BCDMS datasets taken together.
It is seen that the values at NLO and NNLO match within errors with an average value being slightly less for the latter. Note that the values of this parameter at NLO is still compatible with those of the analogous analysis carried out in~\cite{ViMi}.

\vspace{0.5cm}
{\bf Table 2.} Parameter values of the twist-four term in NNLO analysis
\begin{center}
\begin{tabular}{|l||c|c|}
\hline
& &   \\
$x$ & $\tilde h_4(x)$ of $H_2$ $\pm$ stat &  $\tilde h_4(x)$ of $D_2$ $\pm$ stat \\
\hline \hline
0.275& -0.183 $\pm$ 0.020  & -0.197 $\pm$ 0.009 \\
0.35 & -0.149 $\pm$ 0.028  & -0.171 $\pm$ 0.015 \\
0.45 & -0.182 $\pm$ 0.029  & -0.033 $\pm$ 0.031 \\
0.55 & -0.236 $\pm$ 0.052  &  0.142 $\pm$ 0.057 \\
0.65 & -0.180 $\pm$ 0.135  &  0.295 $\pm$ 0.108 \\
0.75 & -0.177 $\pm$ 0.182  &  0.303 $\pm$ 0.158 \\
\hline
\end{tabular}
\end{center}

For illustrative purposes we visualize them in Fig.~1, where fairly good agreement between higher twist corrections obtained at NLO and NNLO is observed, that is not in contradiction with earlier studies (see, for example,~\cite{Alekhin03}).
\begin{figure}[!htb] 
\unitlength=1mm
\vskip -1.5cm
\begin{picture}(0,100)
  \put(0,-5){%
   \psfig{file=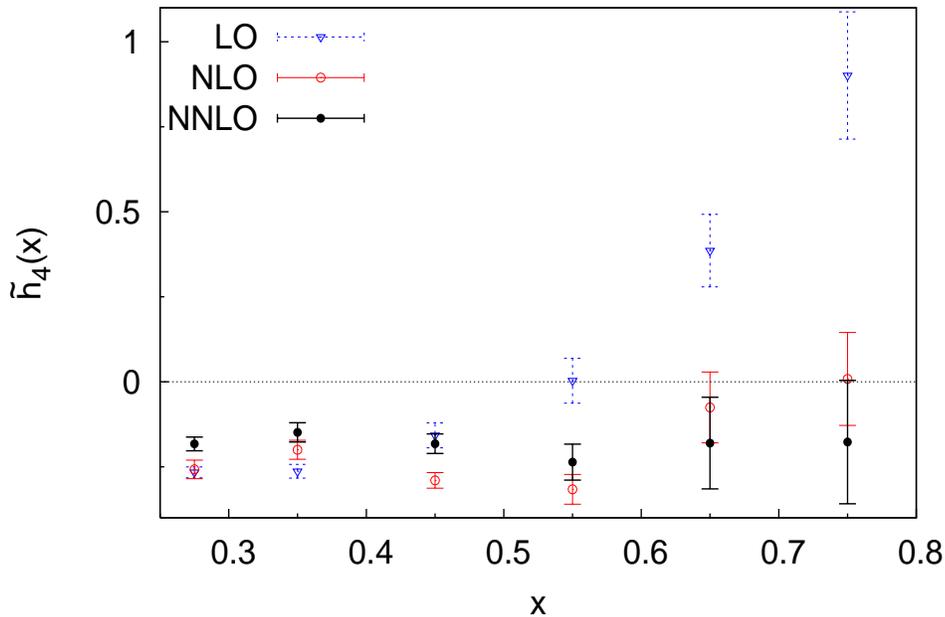,width=130mm,height=95mm}
}
\end{picture}
\vskip 0.2cm
\caption{\sl Comparison of the HTC parameter $\tilde h_4(x)$ obtained at LO, NLO and NNLO in the case of hydrogen data (the bars denote statistical errors).}
\end{figure}

We would like to note that the cut of the BCDMS data, which has increased the
$\as$ values (see Fig.~2 in~\cite{BKKP}) essentially improves agreement between perturbative QCD and experimental data.

\begin{figure}[!htb] 
\unitlength=1mm
\vskip -1.5cm
\begin{picture}(0,100)
\put(0,-5){\psfig{file=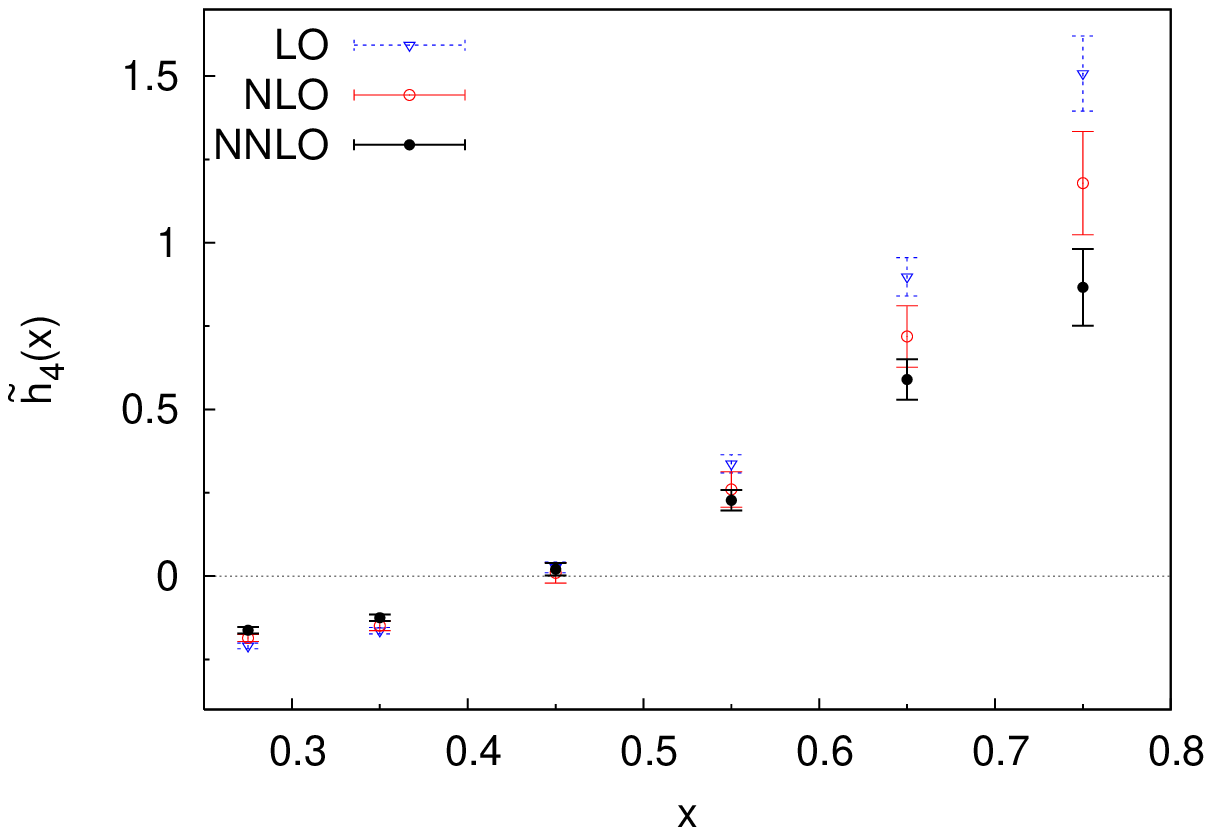,width=130mm,height=95mm}}
\end{picture}
\vskip 0.2cm
\caption{\sl Comparison of the HTC parameter $\tilde h_4(x)$ obtained at LO, NLO and NNLO in the case of hydrogen data within a fixed-flavor-number scheme ($n_f=4$) and with no $y$ cuts imposed on the BCDMS data.}
\end{figure}

Indeed, the HTCs that are nothing else but the difference between
the twist-two approximation (i.e. pure perturbative QCD contribution) and the
experimental data are seen to become considerably smaller at NLO and NNLO levels, as compared to the NLO HT terms obtained in~\cite{ViMi} and also with the results of analysis obtained within a fixed-flavor-number scheme (with a number of flavors fixed to be 4 and no $Y$-cuts imposed on the BCDMS data (see Fig.~2)).

Thus, using the analyses based on the nonsinglet evolution of the SLAC, NMC, BCDMS
and BFP experimental data for SF $F_2$ with no account for the twist-four corrections
and the cut $Q^2 \geq 8$ GeV$^2$ imposed, we obtain (with $\chi^2/DOF=0.93$)
\bea
\as(M_Z^2) ~=~ 0.1167 \pm 0.0008 ~\mbox{(stat)}
\pm 0.0018~\mbox{(syst)}  \pm 0.0007 ~\mbox{(norm)}
\label{NSfin}
\eea
or
\bea
\as(M_Z^2) ~=~ 0.1167 \pm 0.0021~\mbox{(total exp.error)}\,.
\label{NSfin1}
\eea

Upon including the twist-four corrections, and imposing the cut
$Q^2 \geq 1$ GeV$^2$, the following result is found
(with $\chi^2/DOF=0.98$):
\bea
\as(M_Z^2) ~=~ 0.1167 \pm 0.0010 ~\mbox{(stat)}
\pm 0.0020 ~\mbox{(syst)}  \pm 0.0005 ~\mbox{(norm)}
\label{NSfin2}
\eea
or
\bea
\as(M_Z^2) ~=~ 0.1167 \pm 0.0022 ~\mbox{(total exp.error)}
\label{NSfin3}
\eea

\section{Conclusions}

In this work the Jacobi polynomial expansion method developed in~\cite{Barker, Kri, Kri1}
was used to perform analysis of $Q^2$-evolution of DIS structure function $F_2$
by fitting respective fixed-target experimental data that satisfy the cut $x \geq 0.25$.
Based on the results of fitting the value of the QCD coupling constant at the normalization point was evaluated.
Starting with the reanalysis of BCDMS data by cutting off points with large systematic errors it was shown \cite{BKKP}
that the values of $\asMZ$ rise sharply with the cuts on systematics imposed.
The values $\asMZ$ obtained in various fits are in agreement with each other.
An outcome is that quite a similar result for $\asMZ$ was obtained
\cite{BKKP} in the analysis performed over BCDMS data (with the cuts on systematics) and that derived in the analyses done over the data of the rest, thus permitting us to fit available data altogether.

It turns out that for $Q^2 \geq 3$ GeV$^2$ the formul\ae\, of pure perturbative
QCD (i.e. twist-two approximation along with the target mass corrections)
are enough to achieve good agreement with all the data analyzed.
The reference result is then found to be
\be
\as(M_Z^2) = 0.1167 \pm 0.0008 ~\mbox{(stat)}
\pm 0.0018 ~\mbox{(syst)} \pm 0.0007 ~\mbox{(norm)}, \label{re1n} \\
\ee

Upon adding twist-four corrections, fairly good agreement
between QCD (i.e. first two coefficients of Wilson expansion)
and the data starting already at $Q^2 = 1$ GeV$^2$, where the Wilson
expansion begins to be applicable, is observed.
This way we obtain for the coupling constant at $Z$ mass peak:
\be
\as(M_Z^2) = 0.1167 \pm 0.0007~\mbox{(stat)}
\pm 0.0020~\mbox{(syst)} \pm 0.0005~\mbox{(norm)}\,.
\label{re2n} \\
\ee

Note that there also is good agreement with the
analysis~\cite{H1BCDMS} of the combined H1 and BCDMS data, which was published
by H1 collaboration. Our result for $\as(M_Z^2)$
is also compatible with the world average value for the
coupling constant presented in the review ~\cite{Breview},
\footnote{It should be mentioned that this analysis was carried out over
the data coming from the various experiments and in different orders of
perturbation theory, i.e. from NLO up to N$^{3}$LO.}
\be
\as(M_Z^2) = 0.1184 \pm 0.0007.
\label{re2n1}
\ee

Concerning the contributions of higher twist corrections in the present work the well-known $x$-shape of the twist-four corrections while going from intermediate to large values of the Bjorken variable $x$ is well reproduced.

As a next step, we plan to study the DIS experimental data in the non-singlet approximation along with an application of resummation-like
Grunberg effective charge method~\cite{Grunberg} (as it was done in~\cite{Vovk} at the NLO approximation) and the ``frozen''~\cite{frozen,Zotov} and analytic~\cite{SoShi} versions of the strong coupling constant (see~\cite{Zotov,CIKK09,ShiTer} for recent studies in this direction).
While our less immediate goal, a singlet analysis combined with a non-singlet one, is intended to be considered, with the effects of heavy-quark thresholds taken into account, as a follow-up to the previous study~\cite{KKS} that deals with nonsinglet part only.

\vskip 0.5cm
The work was supported by RFBR grant No.10-02-01259-a. 
The work of GP was supported by the grant Ministerio de Ciencia e Inovacion FPA2008-01177.

\end{document}